\documentclass[12pt]{article}
\usepackage[cp1251]{inputenc}
\usepackage[english]{babel}
\usepackage{indentfirst}
\usepackage{misccorr}
\usepackage{fancyhdr}
\usepackage{onlyamsmath}
\usepackage{graphicx}
\usepackage{floatflt}

\title{Probabilistic interpretation of mechanical motion}
\author{Klapchenko V.I., Teslya Y.M.}
\begin{document}
\maketitle

\begin{center}
\emph{Kyiv National University of Construction and Architecture, Kyiv, Ukraine }
\end{center}
\begin{abstract}
Probabilistic approach to the description of translational motion of macrobodies indicates the emergence of additional order effects oriented in the direction of motion of the body
\end{abstract}
\textbf{Introduction.} Today, the applicability of probabilistic description of particles motion in the microsystems (nucleus, atom, molecule), which are the prerogative of quantum mechanics, have become unquestionable. There is no other description for them but probabilistic. Statistical approach is used quite often and with considerable effectiveness while describing the motion of the ensembles of identical particles, molecular systems in particular, obtaining new, so-called statistical laws.

      There is no doubt that the probabilistic approach can be also applied to any macroscopic body that represents itself the ensemble of many constituent particles when it is moving ordinarily along a straight line. The only question is whether such a probabilistic interpretation of mechanical motion will give us any new laws, new ideas? This work is dedicated to the answer to this question.

      We want to express the velocity of body as a whole through internal chaotic motion of its components, but in doing so involve as few details as possible. Therefore, our approach will be significantly different from generally accepted in statistical mechanics – from coordinate-momentum description of behavior of its components with the introduction of distribution functions, etc. \cite{Feynman}. To ensure the problem statement with a simple application of the notion of probability, we will try to introduce the ordered motion of body by two sets of \emph{equivalent random oriented elementary events} \cite{Hudson}.
      
      \emph{Elementary event} is in our consideration the movement of a component of the body with a certain mass of the body $m_{el0}$ in the specified direction with the velocity $V_{el0}$. In fact, the matter is about the \emph{elementary impulse} of a separate element and it becomes a certain unit of measurement of sets of random events. To make the reproduction of this unit single-valued, we agree to determine the mass $m_{el0}$ and the velocity $V_{el0}$ of structural elements at the moment when the bodies are immovable relatively the frame of reference.

      The choice of the same velocity of structural elements of the body $V_{el0}$ is necessary to make all elementary events to be equal. This involves a certain statistical procedure for determining the value $V_{el0}$. For example, such a statistical procedure should be applied twice at the molecular level of body structure: at the averaging over the value of velocity, which leads to the root-mean-square velocity $v_{rms}$, and at the averaging over the value of projection of velocity ($0 - v_{rms}$), which gives

\begin{equation}
{V_{el0}} = \sqrt {\frac{{v_{rms}^2 + {0^2}}}{2}}  = \frac{{{v_{rms}}}}{{\sqrt 2 }}.
\label{eq:1}
\end{equation}

Although such an averaging is quite rough, it is not so important for the selected approach. However, (1) shows that, for the velocity of elements, if necessary, can be chosen the characteristic velocity of molecules which is the closest by value -- the most probable velocity $v_{mp}$.

\begin{figure}\center
\includegraphics{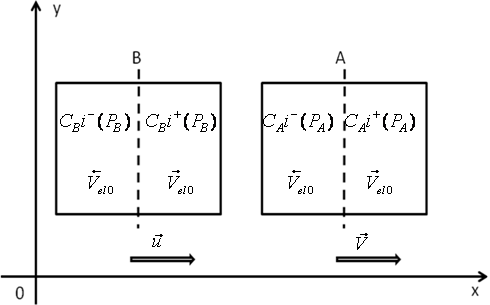}
\caption{Schematic representation of translational motion of two bodies $A$ and $B$. Two sets of events, oriented along the $x$ axis in opposite directions are shown.}
\label{fig:1}
\end{figure}

      \textbf{Examination procedure.} Fig.\ref{fig:1} presents a straight-line uniform motion of two bodies $A$ and $B$ along the positive direction of axis $x$ with velocities $V$ and $u$ by two sets oriented in opposite directions of random elementary events.
       
      For example, the set of oriented events in the direction towards increasing of the axis $x$ (+) for the body $A$ is designated in the figure as the product of $C_A$  and $i^{+}(p_A)$, where $C_A=\dfrac{m_A}{m_{el0}}$ -- mass multiplier of the set of events, $i^+ (p_A)$ -- relative potency of the set of these events. Mass factors $C_A$ and $C_B$ are proportional to the real masses of bodies, and the value of these factors depends on the level of bodies crushing into its component elements. Then the value of velocity of the body can be determined by the probabilities of relevant events in this way. We introduce the value of probability of body motion toward the growth of the axis $x$. For example, for the body $A$

\begin{equation}       
p_A = \frac{C_A i^ + (p_A)}{C_A i^ + (p_A) + C_A i^-(p_A)}=\frac{i^ + (p_A)}{i^ + (p_A) + i^ - (p_A)}.
\label{eq:2}
\end{equation}

The probability of motion in the opposite direction will be respectively equal to $(1-p_A)$. Note that the total number of events, recorded as the sum in the denominator, need not be constant. Then the velocity of the body’s motion will be determined not by the difference of events, but only by the relative difference of events, that is through the probabilities of motion. For example, for the body $A$

\begin{equation}  
V = {p_A}{V_{el0}} - (1 - {p_A}){V_{el0}} = (2{p_A} - 1){V_{el0}}.
\label{eq:3a}
\end{equation}

A similar formula is valid for the body $B$:

\begin{equation}  
u = (2{p_B} - 1){V_{el0}}.
\label{eq:3b}
\end{equation}

      With values of probabilities $\dfrac{1}{2}$ the velocity of the bodies is equal zero that is reasonable and logical. However, so far the formula (\ref{eq:3a}) and (\ref{eq:3b}) do not contribute anything fundamentally new to our understanding of mechanical motion; they can be obtained by direct application of the law of momentum conservation. And in future we must note two circumstances. First. In an imaginary crushing of  bodies to the molecular level the proposed approach can be applied only in cases when the velocity of  bodies is smaller than the velocity of molecules, that is formula (\ref{eq:3a}) and (\ref{eq:3b}) have a limit $u$, $V < V_{el}$. Second. Formula (\ref{eq:3a}) and (\ref{eq:3b}) express only the probabilistic interpretation of mechanical motion, and do not indicate a direct cause-effect relation. The real connection is the following: external reasons for the state of the body (velocity of the elements $V_{el0}$) and the state of motion (or velocity $V$ or $u$), define the potencies of the sets of oriented events and, eventually, the probability of motion $p$. And not vice versa.

      \textbf{Relativistic motions.} If bodies $A$ and $B$ are moving at velocities approaching the speed of light, the statistical review can be applied only at the deepest level of ``crushing''  of bodies on the components. It's the level of structure, which we maybe have not yet achieved, not only experimentally but also theoretically. But at this level the component elements of bodies can’t have velocities which are less than the speed of light $c$ (otherwise, for some unknown reasons the applicability of the probabilistic approach is under a ban). However, they can’t also have the velocities greater than the speed of light $c$ (in accordance with Einstein's postulate of invariance of the speed of light in vacuum). The only thing remains:  postulate the values of velocity of the elements of the body structure by the analogy with the principle of invariance:

\begin{equation} 
V_{el0} = c.
\label{eq:4}
\end{equation}	

      Then the velocity of bodies in the relativistic case will be determined, instead of (\ref{eq:3a}) and (\ref{eq:3b}), through the speed of light in vacuum:
      
\begin{equation} 
V = (2{p_A} - 1)c,
\label{eq:5}
\end{equation}

\begin{equation} 
u = (2{p_B} - 1)c.
\label{eq:6}
\end{equation}

That is, the velocity asymptotically approaches C in all material bodies with non-zero rest mass, and only the photon (electromagnetic wave) in a vacuum will have this value of velocity (in the photon $p$ is $0$ or $1$, that is all the component elements are moving in one direction and at the same velocity).

We shall put now a question about the determination of velocity of the body $A$ relatively the body $B$. To do this, it is necessary to use another postulate -- the equality of inertial reference frames. In our approach this is equivalent to that the relative velocity of bodies must be determined by (\ref{eq:5}) or (\ref{eq:6}), and the probability of the relative motion of $p_{AB}$ can be set by the formula similar to (\ref{eq:2}). At this, the events, which determine the relative movement of two bodies, are composed of two independent events, the probability of which is defined by product of the probabilities of two independent events \cite{Hudson}.

According to Fig. 1, the total number of added events is $4$. Two of them are ``empty'', that means those that do not lead to the relative motion of bodies: $p_A p_B$ and $(1-p_A)(1-p_B)$. They do not participate in determining the probabilities relatively the motion. The other two form this probability: the probability of relative displacement of the body $A$ to the right relatively the body $B$ defines the product $p_A(1-p_B)$and to the left -- the product $(1-p_A)p_B$. Thus,

\begin{equation} 
{p_{AB}} = \frac{{{p_A}(1 - {p_B})}}{{{p_A}(1 - {p_B}) + {p_B}(1 - {p_A})}}.
\label{eq:7}
\end{equation}

is the relative probability of relative displacement of two bodies. Denoting the relative velocity of the body $A$ relatively the body $B$ as $v$, we obtain

\begin{equation} 
v = (2{p_{AB}} - 1)c.
\label{eq:8}
\end{equation}

Expressing from (\ref{eq:5}) and (\ref{eq:6}) the probabilities $p_A$ and $p_B$

\[{p_A} = \frac{{V + c}}{{2c}};{p_B} = \frac{{u + c}}{{2c}},\]

from (\ref{eq:7}) and (\ref{eq:8}) we get:

\begin{equation} 
v = \left( {\frac{{2(c + V)(c - u)}}{{(c + V)(c - u) + (c + u)(c - V)}} - 1} \right)c = \frac{{V - u}}{{1 - \frac{{Vu}}{{{c^2}}}}}.
\label{eq:9}
\end{equation}

We are more accustomed to another form (\ref{eq:9}). If the body $B$ is considered as  moving reference system (that’s why its velocity is indicated by the generally accepted letter $u$), then the velocity $v$ of the body $A$ relatively the body $B$ is its velocity in the moving frame of reference. Expressing from (\ref{eq:9}) the velocity $V$ of the body $A$ relatively the fixed frame of reference, we will have a relativistic velocity addition formula:

\begin{equation} 
V = \frac{{v + u}}{{1 + \frac{{vu}}{c^2}}}.
\label{eq:10}
\end{equation}

This result is expected \cite{Pauli}, and only proves that the probabilistic approach to the description of mechanical motion is applied by us rightly.
      
\textbf{Relative potency of set of oriented events.} In our consideration, the ratio of potency of events oriented in a certain direction of events forms for each body the relative probabilities of motion, which in turn determine the values of velocities of directional movement. Actually, everything happens vice versa -- the values of the velocities designated to the bodies are determined by relevant statistical values. It should be remembered, though while applying the probabilistic approach, only theirs mathematical relation is important:
      
\begin{equation}
\begin{split} 
&\frac{{{C_A}{i^ + }({p_A})}}{{{C_A}{i^ - }({p_A})}} = \frac{{{i^ + }({p_A})}}{{{i^ - }({p_A})}} = \frac{{{p_A}}}{{1 - {p_A}}},  \\ 
&\frac{{{C_B}{i^ + }({p_B})}}{{{C_B}{i^ - }({p_B})}} = \frac{{{i^ + }({p_B})}}{{{i^ - }({p_B})}} = \frac{{{p_B}}}{{1 - {p_B}}}.
\label{eq:11}
\end{split}
\end{equation}

      That is, the values of $i^+$ and iare the functions only of probability of motion. Then there can be put a question on the way of behavior of their sum
      
\begin{equation}
i(p) = {i^ + }(p) + {i^ - }(p).
 \label{eq:12}
\end{equation}

For immovable bodies $p_A=p_B=0.5$ and, according to (11), $\dfrac{i^+}{i^-}=1$. Then the values of the relative powers of oriented events in the immovable bodies can be accepted as equal $0.5$, so

\begin{equation}
i(0.5) = {i^ + }(0.5) + {i^ - }(0.5) = 1.
 \label{eq:13}
\end{equation}

      If both bodies are moving with the same non-zero velocities in one direction ($p_{A} =  p_{B}  = p	\ne 0.5$), that is when their relative velocity equals zero, then power of compound events of opposite directions become equal

\begin{equation}
{p_A}(1 - {p_B}) = {p_B}(1 - {p_A}).
 \label{eq:14}
\end{equation}

Hence

\begin{equation}
\frac{{{p_A}}}{{1 - {p_A}}} = \frac{{{p_B}}}{{1 - {p_B}}},\frac{{{i^ + }({p_A})}}{{{i^ - }({p_A})}} = \frac{{{i^ + }({p_B})}}{{{i^ - }({p_B})}},
 \label{eq:15}
\end{equation}

or

\begin{equation}
{i^ + }({p_A}){i^ - }({p_B}) = {i^ + }({p_B}){i^ - }({p_A}).
 \label{eq:16}
\end{equation}

Expression (\ref{eq:16}) can be repeated many times for any pairs of bodies that have zero relative velocity. Considering that in this case $p_A = p_B = p$, we come to the conclusion that the products of (\ref{eq:16}) should be a universal probability function:
\[{i^ + }(p){i^ - }(p) = f(p).\]

      We know about this universal function only the following. It should not be proportional to the product $p(1–p)$, because otherwise the speed of light becomes reachable. When $p=0.5$ it’s value $0.5 \cdot 0.5=\dfrac{1}{4}>0$. This function should not pass through zero at any point; even at values of p that tend to zero or one. And the most important: at the band edges of determination of probability p both values $i^+$ and $i$ should approach to zero asymptotically. All these conditions can be performed only starting from the dependence 4 on (fig. \ref{fig:2}) that satisfies the condition:

\begin{equation}
{i^ + }(p){i^ - }(p) = const = \frac{1}{4},
\label{eq:17}
\end{equation}

i.e. when this product is invariant. All other dependencies (1-3 fig. \ref{fig:2}) where $i^+$ and $i$ pass through $0$, are unsatisfactory.
     
      Then solving simple system

\begin{equation}\label{eq:18}
\left\{
\begin{array}{rcl}
\dfrac{i^ + }{i^ - } &=& \dfrac{p}{{1 - p}};  \\
i^ +  i^ - &=& \dfrac{1}{4}.\end{array}
\right.
\end{equation}


we obtain 
 
\begin{equation}
\begin{split}
{i^ + } = \frac{1}{2}\sqrt {\frac{p}{{1 - p}}} ,\\
{i^ - } = \frac{1}{2}\sqrt {\frac{{1 - p}}{p}}, 
 \label{eq:19}
 \end{split}
\end{equation}

and for total relative potency of oriented events set
\begin{equation}
i = \frac{1}{{2\sqrt {p(1 - p)} }}.
 \label{eq:20}
\end{equation}

      Let’s express the probability from (\ref{eq:5}) through the movement velocity of body $A$. Then we obtain: 
      
\begin{figure}\center
\includegraphics[width=0.8\textwidth]{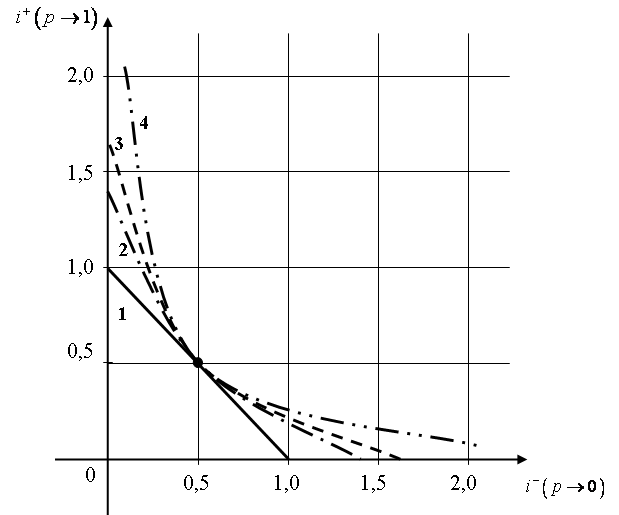}
\caption{To the choice of universal function $i^+ \cdot i^- =f(p)$. Presented field of admissible values of oriented events potency (the first quadrant $i^+ 0 i^-$ ): 1 -- dependency $i^+  i^- = f(p)= p(1-p)$, where $i^+ + i^-=1$; 2, 3 -- intermediate dependences, where $i^+$ or $i:-$ reaches or even passes zero, 4 -- dependency $i^+ \cdot i^- =f(p)=const=\dfrac{1}{4}$ that has a hyperbolic asymptote for $i^+$ (when $p \to 1$) and $i^-$ (when $p \to 0$);}
\label{fig:2}
\end{figure}

\begin{equation}
i = \frac{1}{{\sqrt {1 - \frac{{{V^2}}}{{{c^2}}}} }},
 \label{eq:21}
\end{equation}

that coincides with the multiplier in the Lorentz transformation of coordinates equations \cite{Pauli}. Let’s find out what is the physical meaning of a full set of oriented events of body $A$. To do this, the relative potency of a set (21) must be multiplied by the mass multiplier $C_A$ and by the value of the elementary impulse $m_{el0}V_{el0}$:

\begin{equation}
i{C_A}{m_{el0}}{V_{el0}} = \frac{{{m_0}c}}{{\sqrt {1 - \frac{{{V^2}}}{{{c^2}}}} }} = mc.
 \label{eq:22}
\end{equation}

Where m equals

\begin{equation}
m = \frac{{{m_0}}}{{\sqrt {1 - \frac{{{V^2}}}{{{c^2}}}} }},
 \label{eq:23}
\end{equation}

that is the one of the consequences of the special theory of relativity. By the way, Lorentz was emerged by idea of anisotropy of moving electron mass \cite{Lorentz}. Multiplier (\ref{eq:21}), more precisely its square, determines the ratio of longitudinal electron mass to the transverse mass.

      But in any case we got something new for our understanding of the fundamentals and implications of special relativity. This new comes to the following: with increasing of body velocity, increases the number of events in the body, oriented in the direction of its movement. Then you can calculate the relative share of mainly oriented events (original index of mechanical movement anisotropy):

\begin{equation}
\beta  = \frac{{{i^ + } - {i^ - }}}{{{i^ + } + {i^ - }}} = \frac{\delta }{i} = 2p - 1 = \frac{V}{c},
 \label{eq:24}
\end{equation}

and to find out the physical content of a mainly oriented events set:

\begin{equation}
\delta C_A m_{el0} V_{el0} = \frac{V}{c} \cdot \frac{{m_0 c}}{{\sqrt {1 - \frac{{{V^2}}}{{{c^2}}}} }} = mV.
 \label{eq:25}
\end{equation}

That is, the mainly oriented event set is actually a relativistic momentum. Then expressions (\ref{eq:23}) and (\ref{eq:25}) allow to construct a relativistic invariant

\[(mc)^2 - {(mV)^2} = {({m_0}c)^2},\]
that’s equivalent to the equation ${i^2} - \delta^2 = {({i^ + } + {i^ - })^2} - {({i^ + } - {i^ - })^2} = 1$, or: 

\begin{equation}
4{i^ + }{i^ - } = 1.
\label{eq:26}
\end{equation}

That permits us to name (\ref{eq:17}) also invariant.
      If we use index $\beta$, expressions for potency of mainly oriented events and total oriented events potency will be the shortest:
      
\begin{equation}
\begin{split}
& \delta  = {i^ + } - {i^ - } = \frac{\beta }{\sqrt {1 - \beta ^2}} ;\\ 
& i = {i^ + } + {i^ - } = \frac{1}{\sqrt {1 - \beta ^2}}. 
\label{eq:27}
\end{split}
\end{equation}

\textbf{Transformation of ellipse and ellipsoid events.} Let’s provide the most obvious appearance to all new that we learned about moving bodies. Graphic representations will be complete when besides potency $i$ of set of events (oriented along the direction of movement) there will be also shown potencies of sets of events in the transverse directions $i_y$, $i_z$ , which were not considered earlier. To take into account anisotropy of events, let’s use Lorentz’s \cite{Lorentz} connection between the potency of events along the movement direction and across it, similar to mass ratio that we talked about the formula (\ref{eq:23}):

\begin{equation}
\frac{i}{i_y} = \frac{{{i^ + } + {i^ - }}}{{i_y^ +  + i_y^ - }} = \frac{1}{1 - \beta ^2}.
   \label{eq:28}
\end{equation}

      Then, using the above stated values $i$, $i^+$, $i^-$, connection specified as formula (\ref{eq:28}) and symmetry of transverse to the $x$ axis directions, we obtain:  

\begin{equation}
\begin{split}
& i_y^+  = i_y^ -  = \frac{i_y}{2} = \frac{\sqrt {1 - \beta ^2}}{2};\\
& i{i_y} = 1;  \\ 
& {i^ + } = \frac{i_y}{2(1 - \beta )}; \\
& {i^ - } = \frac{i_y}{2(1 + \beta )}. 
\label{eq:29}
\end{split}
\end{equation}

\begin{figure}\center
\includegraphics[width=0.8\textwidth]{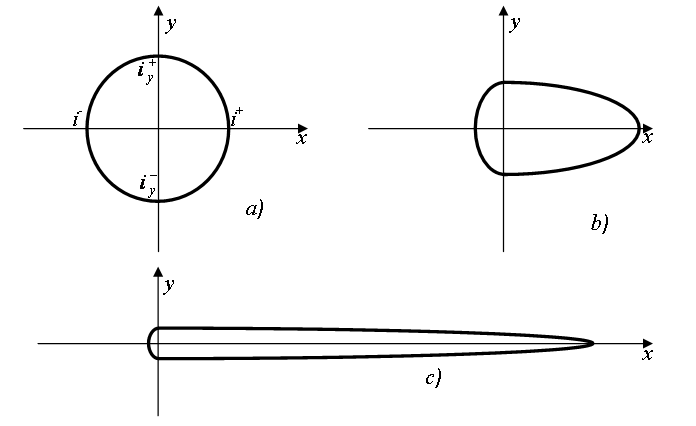}
\caption{Schematic representation of events ellipse transformation.}
\label{fig:3}
\end{figure}

Fig. \ref{fig:3} presents an intersection of plane $x0y$ and ellipsoid of revolution that is constructed with the relative events set potencies along the movement direction ($x$ axis) and across it. Then the imaginary area of the events set transforms from a circle (the intersection of the sphere) with radius of $0.5$ at $\beta=0$ (fig. \ref{fig:3} a), to the distorted ellipse (the intersection of an ellipsoid of revolution around the $x$ axis) at an intermediate value of $\beta$ (fig. \ref{fig:3} b), and even to the prolate distorted ellipse at $\beta$ urging to $1$ (fig. \ref{fig:3} c).

Simply, the transformation of the events set undergo from the sphere stage (immovable body) to a prolate into the thread ellipsoid -- events needle -- when $\beta \to 1$. It is noteworthy that the plane of events set appearances (ellipsoid of revolution intersection plane) remains unchanged:
      
\begin{equation}
s = \pi i{i_y} = \pi  = const.
\label{eq:30}
\end{equation}

      Unlike the intersection area, events ellipsoid volume of a moving body is reducing. Let designate the volume as letter $\Omega$, then for an observer in the immovable frame of reference:

\begin{equation}
\Omega  = \frac{\pi }{6}ii_y^2 = \frac{\pi }{6} \sqrt {1 - \beta^2}. 
\label{eq:31}
\end{equation}

      At the same time for an observer that moves together with the body, area and volume of events ellipsoid in his reference frame degenerates into a circle or sphere with a diameter of $iy$, so: 
\begin{equation}
s' = \pi i_y^2 = \pi \left( {1 - {\beta ^2}} \right);\Omega ' = \frac{\pi}{6}i_{y^3} = \frac{\pi }{6}{\left( {1 - \beta ^2} \right)^{3/2}}.
\label{eq:32}
\end{equation}

Let’s emphasize for the future that the important result is that the ratio of volumes of events ellipsoid to the area of the events ellipse for both observers are the same:

\begin{equation}
\frac{\Omega }{s} = \frac{\pi ii_y^2}{6\pi ii_y} = \frac{1}{6}{\left( {1 - \beta ^2} \right)^{1/2}} ,
\label{eq:33}
\end{equation}

\begin{equation}
\frac{\Omega '}{s'} = \frac{\pi i_y^3}{6\pi i_y^2} = \frac{1}{6}{\left( {1 - \beta ^2} \right)^{1/2}}.
\label{eq:34}
\end{equation}

That is, we are dealing with one more relativistic invariant physical content of which we will determine later.

\textbf{Conclusions.}

\begin{itemize}
	\item The applicability of the probabilistic approach to the consideration of  bodies translational motion. Accuracy and reasonableness of the obtained results is confirmed by complete coincidence with results of other trials, including the relativistic field;
	 \item the use of the probabilistic approach leads to new concepts in the description of motion that couldn’t be obtained in other approaches. In particular, increasing the velocity of the body leads to an increase in the number of body events oriented in the direction of movement. Anisotropy of mechanical movement is most evident when the velocity of the body approaches to the speed of light, becoming a peculiar needle of events.
\end{itemize}

\emergencystretch=3pt

\end{document}